\documentclass{mn2e}
\usepackage{graphics}
\usepackage{epsfig}

\def\etal{et\thinspace al.\ }                               

\def\beq{\begin{equation}}                          
\def\eeq{\end{equation}}                              
\def\beqa{\begin{eqnarray}}                         
\def\eeqa{\end{eqnarray}}                             
\def\beqan{\begin{eqnarray*}}                      
\def\eeqan{\end{eqnarray*}}                          

\newbox\grsign \setbox\grsign=\hbox{$>$} \newdimen\grdimen \grdimen=\ht\grsign
\newbox\simlessbox \newbox\simgreatbox
\setbox\simgreatbox=\hbox{\raise.5ex\hbox{$>$}\llap
     {\lower.5ex\hbox{$\sim$}}}\ht1=\grdimen\dp1=0pt
\setbox\simlessbox=\hbox{\raise.5ex\hbox{$<$}\llap
     {\lower.5ex\hbox{$\sim$}}}\ht2=\grdimen\dp2=0pt

\title[Emission Line Properties of Seyfert 2 Nuclei]
   {Emission Line Properties of Seyfert 2 Nuclei}

\author[Gu, Melnick, Cid Fernandes, Kunth, Terlevich, Terlevich]
       {Q. Gu$^{1}$\thanks{E-mail: qsgu@nju.edu.cn},
        J. Melnick$^{2}$\thanks{E-mail: jmelnick@eso.org},
        R. Cid Fernandes$^{3}$\thanks{E-mail: cid@astro.ufsc.br},
        D. Kunth$^{4}$\thanks{E-mail: kunth@iap.fr},
        E. Terlevich$^{5}$\thanks{E-mail: eterlevi@inaoep.mx. Visiting fellow, Institute of Astronomy, Cambridge, U.K.},
    \newauthor
        R. Terlevich$^{5}$\thanks{E-mail: rjt@inaoep.mx. Visiting fellow, Institute of Astronomy, Cambridge, U.K.}
\newauthor
\\
   $^{1}$Department of Astronomy, Nanjing University, Nanjing 210093, P. R. China\\
   $^{2}$European Southern Observatory, Alonso de Cordova 3107, Santiago, Chile\\
   $^{3}$Departamento de F\'{\i}sica - CFM - Universidade Federal de Santa Catarina, PO Box 476,
         Florian\'opolis 88040-900, SC, Brazil\\
   $^{4}$Institut d'Astrophysique de Paris, 98bis Boulevard Arago, 75014 Paris, France\\
   $^{5}$Instituto Nacional de Astrof\'\i sica, Optica y Electr\'onica,
AP 51 y 216, 72000, Puebla, M{\'e}xico}

\begin{document}

\maketitle

\begin{abstract}

This is the third paper of a series devoted to the study of the
global properties of Joguet's sample of 79 nearby galaxies
observable from the southern hemisphere, of which 65 are Seyfert 2
galaxies.  We use the population synthesis models of Paper~II to
derive `pure' emission-line spectra for the Seyfert 2's in the
sample, and thus explore the statistical properties of the nuclear
nebular components and their relation to the stellar populations.
We find that the emission line clouds suffer substantially more
extinction than the starlight, and confirm the correlations
between stellar and nebular velocity dispersions and between
emission line luminosity and velocity dispersions, although with
substantial scatter. Nuclear luminosities correlate with stellar
velocity dispersions, but Seyferts with conspicuous star-forming
activity deviate systematically towards higher luminosities.
Removing the contribution of young stars to the optical continuum
produces a tighter and steeper relation, $L \propto
\sigma_\star^4$, consistent with the Faber-Jackson law.

Emission line ratios indicative of the gas excitation such as
[OIII]/H$\beta$ and [OIII]/[OII] are statistically smaller for
Seyferts with significant star-formation, implying that ionization
by massive stars is responsible for a substantial, and sometimes
even a dominant, fraction of the H$\beta$ and [OII] fluxes. We use
our models  to constrain the maximum fraction of the ionizing
power that can be generated by a hidden AGN. We correlate this
fraction with classical indicators of AGN photo-ionization: X-ray
luminosity and nebular excitation, but find no significant
correlations. Thus, while there is a strong contribution of
starbursts to the excitation of the nuclear nebular emission in
low-luminosity Seyferts, the contribution of the hidden AGN
remains elusive even in hard X-rays.
\end{abstract}


\section{Introduction}

 The emission-line fluxes and profiles of active galactic nuclei
 (AGNs) carry, at least in principle, all the  information
 necessary  to study the properties of both the broad-line (BLR)
 and the narrow-line (NLR) regions, such as ionization and
 excitation, extinction, metallicity, kinematics, and  AGN power
 (Heckman et al., 1981; Osterbrock \& Shuder 1982; Wilson \& Nath
 1990; Whittle 1992a,b,c; Nelson \& Whittle 1996; Ho et al. 2003;
 Hao et al. 2005).

 An important and unavoidable issue in the study of the
 emission-line spectra, however, is contamination by the
 underlying stellar populations, particularly for the Balmer
 lines. Historically, McCall, Rybski, \& Shields (1985)
 recommended an empirical absorption correction of 2 \AA \ to the
 equivalent widths (EWs) for H$\alpha$, H$\beta$ and H$\gamma$,
 which, of course, is rather coarse. Nowadays, with available high
 spectral resolution stellar population synthesis models, such as
 Bruzual \& Charlot (2003), or Gonz\'alez Delgado et al (2005) for
 example,  it is possible to fit the observed absorption lines and
 emission-line-free continua simultaneously and thus to obtain
 the pure emission-line spectra after subtraction of the
 synthesized stellar  components (Tremonti 2003; Kauffmann et al.
 2003a,b; Hao et al. 2005; Cid Fernandes et al. 2005).

 This is the purpose of the present paper, the third  of a series
 devoted to the study of the stellar populations in the nuclear
 regions of Seyfert 2 galaxies of Joguet's sample of nearby
 galaxies. The sample comprises 79 southern galaxies
 ($\delta<15^{\deg}$) classified as Seyfert~2's with redshifts
 lower than z=0.017.  A full description of the sample is
 presented in Joguet et al.  (2001; Paper I).   After a  careful
 inspection of the data set including a revision of the data in
 the literature we found in Paper~II  that of the 79 sample
 galaxies, one is a type 1 Seyfert, 65 are bona-fide Seyfert 2's,
 4 are  LINER's, 4 are starburst/HII nuclei, and the remaining 5
 are normal galaxies showing no nuclear emission lines.

 We found in Cid Fernandes et al. (2004; Paper~II) that the nuclear
 regions of the true Seyfert galaxies in the sample present
 remarkably heterogeneous star formation histories where young
 starbursts, intermediate age, and old stellar populations all
 appear in significant, but widely variable proportions.  We also
 found that a significant fraction of the nuclei  show a strong
 featureless continuum (FC) component, but that this component is
 not always an indication of a hidden AGN; it could also betray
 the presence of young dusty starbursts 
 (Storchi-Bergmann et al. 2000; Gonzalez Delgado et al.
 2001 and Cid Fernandes et al. 2001).

 The present paper, the last of the series dealing with Joguet's
 (2001) sample, is devoted to the study of the emission line
 properties of the 65 Seyfert~2 nuclei in the sample and
 their relation with the stellar components. In Section 2 we
 measure the emission line fluxes from the pure emission line
 spectra and derive the nebular extinction from Balmer decrements.
 Section 3 presents the relation between stellar and nebular
 velocity dispersion and Faber-Jackson relation for Seyfert 2 galaxies.
 The implications and discussion of the results are given in Section 4
 and conclusions in Section 5.

\section{The data}

 By subtracting the synthetic stellar components we were able in
 Paper~II to obtain remarkably clean pure emission line spectra.
 Examples of these `residual' spectra were presented in
 Figures~2-10 of Paper~II. Using standard tools within IRAF
 \footnote{IRAF is distributed by the National Optical Astronomy
 Observatories, which is operated by the Association of
 Universities for Research in Astronomy, Inc., under cooperative
 agreement  with the National Science Foundation.} we measured the
 strengths of the most prominent emission features in the residual
 spectra, including line fluxes and measurement errors, line
 widths, and equivalent widths using the continua from the
 synthetic spectra.  The results are presented in Table~1
 including estimates for the measurement errors derived from the
 photon statistics. Since many of our nuclei show faint, broad
 non-Gaussian wings, the widths were measured using two methods:
 fitting single Gaussians to the [OIII]$\lambda5007$ line, and
 manually measuring the widths at half-intensity. The two
 measurements agree to better than 10\% so it is safe to interpret
 the line widths as true velocity dispersions.  The velocity
 dispersion from the Gaussian fits and the associated errors
 obtained using ({\it ngaussfit} in IRAF) are given in the Table 1 
 where for completeness we also tabulate the stellar velocity
 dispersions ($\sigma_{\star}$) from Paper~II. While our
 population synthesis code (Paper~II) does not provide estimates
 of the internal errors associated with the stellar velocity
 dispersions, a comparison with published results from other
 groups using different techniques, indicates that the typical
 errors in our values are $\sim 20 km/s$. The absorption-corrected
 hard X-ray luminosities (2-10 keV) collected from the literature
 and the references to the corresponding publications are given in
 the last two columns of Table~1. The lower-limits correspond to
 Compton-thick sources  -- column densities larger than 10$^{24}
 cm^{-2}$ -- for which the direct hard X-ray continua are
 completely absorbed.

\begin{table*}
\begin{center}
\scriptsize \caption{Emission line properties of Seyfert~2
galaxies}
\begin{tabular}{lllllllllll}
\hline\hline
Galaxy  & [OII]$^a$    & H$\delta^a$ & H$\gamma^a$ & HeII$^a$  & H$\beta^a$   & O[III]$^a$      & $\rm \sigma_{\rm [OIII]}$ &$\rm \sigma_*$ & L$_{X}$ & Ref \\
 \tiny      & 3727 & 4101 &  4340 & 4686  & 4861 & 5007  & km s$^{-1}$   & km s$^{-1}$  &  erg s$^{-1}$ & \\
    \hline
 ESO 103-G35    &              135.3 $\pm$  8.5  &   10.1 $\pm$  2.4  &  18.2 $\pm$  3.1 &    4.8 $\pm$  1.5  &   44.8 $\pm$  5.2 &   430.3  $\pm$ 15.6    &  259 $\pm$ 8.1 & 114 & 42.79  & 1 \\
 ESO 104-G11    &               45.7 $\pm$  5.0  &    6.9 $\pm$  1.8  &  12.3 $\pm$  2.5 &    7.6 $\pm$  1.6  &   26.7 $\pm$  3.6 &   279.8  $\pm$ 12.4    &  177 $\pm$ 2.8 & 130 &        & \\
 ESO 137-G34    &              719.0 $\pm$ 19.2  &   66.3 $\pm$  5.7  & 107.0 $\pm$  7.4 &   55.3 $\pm$  5.1  &  240.0 $\pm$ 11.1 &  2541.0  $\pm$ 36.0    &  239 $\pm$ 3.8 & 133 &        & \\
 ESO 138-G01    &              902.3 $\pm$ 21.5  &  107.5 $\pm$  7.4  & 197.0 $\pm$ 10.1 &  120.0 $\pm$  7.9  &  433.0 $\pm$ 15.1 &  3693.0  $\pm$ 44.5    &  135 $\pm$ 1.8 &  80 & $>$41.46$^b$& 2 \\
 ESO 269-G12    &               45.0 $\pm$  4.9  &    3.3 $\pm$  1.3  &   4.8 $\pm$  1.6 &                    &    7.8 $\pm$  1.9 &    58.6  $\pm$  5.5    &  150$\pm$ 11.3 & 161 &       &\\ \\
 ESO 323-G32    &               80.1 $\pm$  6.3  &   10.7 $\pm$  2.8  &  15.3 $\pm$  2.6 &   13.7 $\pm$  2.4  &   36.8 $\pm$  4.1 &   460.6  $\pm$ 15.6    &  132 $\pm$ 1.6  & 131 &        &\\
 ESO 362-G08    &               48.0 $\pm$  5.1  &   15.6 $\pm$  2.8  &  22.5 $\pm$  2.6 &   13.3 $\pm$  3.2  &   38.4 $\pm$  4.4 &   286.2  $\pm$ 12.2    &  161 $\pm$ 2.3 & 154 &        & \\
 ESO 373-G29    &              186.6 $\pm$  9.7  &   40.0 $\pm$  4.4  &  65.0 $\pm$  5.7 &   25.1 $\pm$  3.6  &  142.3 $\pm$  8.5 &   751.0  $\pm$ 19.9    &   78 $\pm$ 1.3 &  92 &        &\\
 ESO 381-G08    &              444.7 $\pm$ 15.2  &   62.8 $\pm$  5.9  & 127.5 $\pm$  8.5 &   39.5 $\pm$  4.8  &  301.0 $\pm$ 13.1 &  1942.0  $\pm$ 32.7    &  206 $\pm$ 3.8 & 100 &        & \\
 ESO 383-G18    &               80.0 $\pm$  6.3  &    9.4 $\pm$  2.2  &  18.2 $\pm$  3.0 &                   &   42.5 $\pm$  4.7 &   360.9  $\pm$ 13.6    &  105 $\pm$ 1.6 &  92 &        & \\ \\
 ESO 428-G14    &              910.5 $\pm$ 21.7  &   93.1 $\pm$  7.4  & 187.4 $\pm$  9.8 &  126.8 $\pm$  7.8  &  397.0 $\pm$ 14.6 &  5198.0  $\pm$ 52.3    &  187 $\pm$ 2.2 & 120 & $>$40.64$^b$ & 3\\
 ESO 434-G40    &              118.8 $\pm$  7.9  &   28.9 $\pm$  3.6  &  37.6 $\pm$  4.1 &   32.7 $\pm$  3.8  &   84.5 $\pm$  6.8 &   913.4  $\pm$ 22.0    &  106 $\pm$ 1.4 & 145 & 43.06  & 1 \\
 Fairall 334      &               26.9 $\pm$  3.7  &    3.5 $\pm$  1.3  &   6.4 $\pm$  1.7 &        $\pm$       &   14.3 $\pm$  2.7 &    85.6  $\pm$  6.8    &  130 $\pm$ 2.8 & 104 & 42.95  & 4 \\
 Fairall 341      &               40.0 $\pm$  4.3  &    7.7 $\pm$  1.6  &   9.7 $\pm$  2.2 &    8.8 $\pm$  2.1  &   25.1 $\pm$  3.4 &   279.2  $\pm$ 12.1    &  120 $\pm$ 1.6 & 122 & 43.03  & 4 \\
   IC 1657      &               34.3 $\pm$  4.2  &                    &   4.7 $\pm$  1.5 &        $\pm$       &    9.7 $\pm$  2.2 &    25.7  $\pm$  3.6    &  148 $\pm$ 6.5 & 143 &        & \\ \\
   IC 2560      &              175.9 $\pm$  9.6  &   40.5 $\pm$  4.5  &  61.9 $\pm$  5.6 &   49.2 $\pm$  5.0  &  135.0 $\pm$  8.0 &  1250.0  $\pm$ 25.8    &  135 $\pm$ 1.6 & 144 & $>$40.94$^b$ & 5 \\
   IC 5063      &              247.4 $\pm$ 11.4  &   22.7 $\pm$  3.5  &  41.5 $\pm$  4.7 &   15.6 $\pm$  2.6  &  106.0 $\pm$  7.5 &   911.4  $\pm$ 22.2    &  183 $\pm$ 2.7 & 182 & 42.87  & 1\\
   IC 5135      &              305.2 $\pm$ 12.4  &   34.6 $\pm$  4.7  &  80.0 $\pm$  6.7 &   32.8 $\pm$  4.0  &  191.6 $\pm$ 10.5 &  1108.0  $\pm$ 24.6    &  316 $\pm$ 8.5 & 143 & $>$41.40$^b$ & 6\\
 F11215-2806&               51.8 $\pm$  5.2  &    4.6 $\pm$  1.7  &   8.1 $\pm$  2.1 &    6.5 $\pm$  1.9  &   22.5 $\pm$  3.6 &   243.2  $\pm$ 11.9    &  184 $\pm$ 2.4 &  98 &        &   \\
 M+01-27-20 &               65.4 $\pm$  5.7  &    9.2 $\pm$  2.1  &  15.7 $\pm$  2.8 &    6.4 $\pm$  1.7  &   39.9 $\pm$  4.5 &   157.0  $\pm$  9.0    &  103 $\pm$ 2.1 &  94 &        &  \\ \\
 M-03-34-64 &              310.3 $\pm$ 13.1  &   62.8 $\pm$  6.1  & 130.4 $\pm$  8.7 &  145.3 $\pm$  8.2  &  320.9 $\pm$ 14.0 &  4434.0  $\pm$ 47.2    &  461 $\pm$ 7.6 & 155 & 42.53  & 1\\
   Mrk 897      &              101.7 $\pm$  7.1  &   23.6 $\pm$  3.7  &  44.9 $\pm$ 4.7  &                    &  103.0 $\pm$ 7.1  &    55.8  $\pm$  5.6    &  154 $\pm$ 34.5 & 133 &        &  \\
   Mrk 1210     &              212.1 $\pm$ 10.8  &   55.1 $\pm$  6.0  & 131.9 $\pm$ 8.7  &   54.3 $\pm$ 5.8   &  282.0 $\pm$ 13.0 &  2847.0  $\pm$ 40.9    &  275 $\pm$ 3.9 & 114 & $>$41.96$^b$ & 1\\
   Mrk 1370     &               49.5 $\pm$ 5.1   &    5.1 $\pm$  1.5  &   8.5 $\pm$ 1.9  &                    &   19.6 $\pm$  3.2 &   128.8  $\pm$  8.4    &  188 $\pm$ 3.9  &  86 &        &  \\
   NGC 424      &              455.2 $\pm$15.7   &  147.2 $\pm$  9.1  & 288.8 $\pm$ 12.9 &  204.1 $\pm$ 10.8  &  766.0 $\pm$ 21.1 &  4267.0  $\pm$ 47.1    &  298 $\pm$ 5.3 & 143 & $>$41.50$^b$ & 2\\ \\
   NGC 788      &               99.2 $\pm$ 7.1   &   10.4 $\pm$  2.3  &  17.3 $\pm$ 2.9  &    8.8 $\pm$ 2.0   &   31.6 $\pm$  4.0 &   364.8  $\pm$ 13.7    &  101 $\pm$ 1.5 & 163 &        &  \\
   NGC 1068     &             3379.0 $\pm$ 43.0  &  744.8 $\pm$  20.5 &1106.0 $\pm$28.3  & 1102.0 $\pm$25.9   & 3194.0 $\pm$43.3  & 3.558e+04$\pm$135.4    &  561 $\pm$ 9.0  & 144 & $>$40.98$^b$ & 1\\
   NGC 1125     &               86.3 $\pm$ 6.7   &    7.9 $\pm$  2.0  &  16.0 $\pm$ 2.8  &    6.3 $\pm$ 1.8   &   35.1 $\pm$4.3   &   227.9  $\pm$ 11.3    &  186 $\pm$ 3.2 & 105 &        &  \\
   NGC 1667     &               38.7 $\pm$ 4.6   &                    &   7.8 $\pm$ 1.8  &                    &   10.5 $\pm$2.2   &    91.3  $\pm$  6.9    &  174 $\pm$ 3.1 & 149 & $>$42.34$^b$ & 1 \\
   NGC 1672     &               97.8 $\pm$ 7.2   &   17.9 $\pm$  3.4  &  46.8 $\pm$ 4.8  &                    &  111.1 $\pm$7.3   &    83.3  $\pm$  6.7    &  129 $\pm$ 42.3 &  97 & 41.46  & 1\\ \\
   NGC 2110     &              944.5 $\pm$ 22.3  &   55.6 $\pm$  6.5  &  98.5 $\pm$ 7.5  &   19.2 $\pm$ 3.0   &  167.1 $\pm$9.5   &   751.9  $\pm$ 20.7    &  199 $\pm$ 7.1 & 242 & 42.57  & 1\\
   NGC 2979     &               34.6 $\pm$ 4.4   &    4.8 $\pm$  1.8  &   7.5 $\pm$ 2.0  &                    &   11.6 $\pm$  2.4 &   109.7  $\pm$  7.6    &  170 $\pm$ 2.9  & 112 &        & \\
   NGC 2992     &              125.6 $\pm$ 8.1   &   13.2 $\pm$  2.8  &  24.4 $\pm$ 3.9  &   10.7 $\pm$ 2.2   &   63.8 $\pm$  6.2 &   519.9  $\pm$ 17.4    &  164 $\pm$ 2.2 & 172 & 41.71  & 1\\
   NGC 3035     &               37.4 $\pm$ 4.3   &                    &   7.8 $\pm$ 1.9  &                    &   19.7 $\pm$  3.6 &   171.2  $\pm$  9.5    &  121 $\pm$ 1.9  & 161 &        & \\
   NGC 3081     &              288.8 $\pm$ 12.2  &   46.6 $\pm$  4.8  &  78.0 $\pm$ 6.4  &   75.1 $\pm$ 6.0   &  172.3 $\pm$ 9.6  &  2109.0  $\pm$ 33.6    &   93 $\pm$ 1.4 & 134 & 41.91 & 1\\ \\
   NGC 3281     &               25.7 $\pm$ 3.6   &    5.1 $\pm$  1.5  &   6.4 $\pm$ 1.8  &    5.4 $\pm$ 1.6   &   17.4 $\pm$ 2.9  &   129.4  $\pm$  8.2    &  141 $\pm$ 3.2 & 160 & 42.79 & 1\\
   NGC 3362     &               57.9 $\pm$ 5.6   &    6.0 $\pm$  1.7  &  10.4 $\pm$ 2.5  &    6.4 $\pm$ 1.9   &   30.5 $\pm$ 3.9  &   237.2  $\pm$ 11.5    &  179 $\pm$ 3.8 & 104 &       & \\
   NGC 3393     &              568.4 $\pm$ 17.3  &   72.1 $\pm$  6.0  & 122.3 $\pm$ 8.1  &   76.9 $\pm$ 6.1   &  270.5 $\pm$ 12.0 &  2680.0  $\pm$ 37.9    &  184 $\pm$ 2.4 & 157 & $>$41.08$^b$ & 1\\
   NGC 3660     &               23.9 $\pm$ 3.5   &    5.5 $\pm$  1.6  &  13.2 $\pm$ 2.9  &    7.1 $\pm$ 1.6   &   39.8 $\pm$ 5.2  &   105.2  $\pm$  7.4    &   90 $\pm$ 6.3 &  95 &       & \\
   NGC 4388     &              501.9 $\pm$ 16.1  &   30.6 $\pm$  3.9  &  56.4 $\pm$ 5.5  &   31.9 $\pm$  4.0  &  145.1 $\pm$ 8.8  &  1621.0  $\pm$ 29.6    &  152 $\pm$ 2.0 & 111 & 42.76 & 1\\ \\
   NGC 4507     &             1022.0 $\pm$ 23.5  &  134.6 $\pm$  8.7  & 262.0 $\pm$ 12.3 &  100.4 $\pm$ 7.5   &  578.8 $\pm$ 18.5 &  4455.0  $\pm$ 51.7    &  213 $\pm$ 3.6 & 144 & 43.28 & 1\\
   NGC 4903     &               52.3 $\pm$ 4.8   &    8.2 $\pm$  2.2  &  10.3 $\pm$ 2.4  &    8.8 $\pm$  1.8  &   28.7 $\pm$ 3.7  &   262.0  $\pm$ 11.6    &  158 $\pm$ 1.8 & 200 &       & \\
   NGC 4939     &              250.5 $\pm$ 11.2  &   39.8 $\pm$  4.4  &  61.0 $\pm$ 5.6  &   52.1 $\pm$  4.9  &  138.5 $\pm$ 8.6  &  1599.0  $\pm$ 29.0    &  163 $\pm$ 1.8 & 155 & 41.96 & 1\\
   NGC 4941     &              276.5 $\pm$ 11.9  &   27.7 $\pm$  3.7  &  47.5 $\pm$ 5.0  &   36.7 $\pm$  3.8  &  114.4 $\pm$ 7.8  &  1426.0  $\pm$ 27.9    &  119 $\pm$ 2.2 &  98 & 40.90 & 1\\
   NGC 4968     &               40.5 $\pm$ 4.7   &   10.3 $\pm$  2.3  &  14.5 $\pm$ 2.7  &    9.1 $\pm$  2.3  &   36.7 $\pm$ 4.5  &   434.4  $\pm$ 15.5    &  176 $\pm$ 1.9 & 121 & $>$40.87$^b$ & 1\\
 \hline\hline
\end{tabular}
\end{center}
\label{data}
\end{table*}

\setcounter{table}{0}
\begin{table*}
\begin{center}
\scriptsize \caption{Emission line properties of Seyfert~2
galaxies (continued)}
\begin{tabular}{lllllllllll}
\hline\hline
 Galaxy  & [OII]$^a$    & H$\delta^a$ & H$\gamma^a$ & HeII$^a$  & H$\beta^a$   & O[III]$^a$      & $\rm \sigma_{\rm [OIII]}$ &$\rm \sigma_*$ & L$_{X}$ & Ref \\
        & 3727 & 4101 &  4340 & 4686  & 4861 & 5007  & km s$^{-1}$   &  km s$^{-1}$ & erg s$^{-1}$  &  \\
    \hline
   NGC 5135     &              133.1 $\pm$ 8.3   &   21.0 $\pm$  3.2  &  41.0 $\pm$ 4.5  &   20.6 $\pm$  2.8  &   99.6 $\pm$ 7.3  &   363.6  $\pm$ 14.5    &  155 $\pm$ 4.4 & 143 & $>$40.86$^b$ & 1\\
   NGC 5252     &              186.6 $\pm$ 9.8   &   12.2 $\pm$  2.5  &  21.7 $\pm$ 3.4  &    4.6 $\pm$  1.2  &   40.8 $\pm$ 4.6  &   268.1  $\pm$ 12.0    &  195 $\pm$ 6.9 & 209 & 43.07 & 1\\
   NGC 5427     &               38.0 $\pm$ 4.5   &    7.0 $\pm$  1.9  &   8.3 $\pm$ 1.9  &    3.9 $\pm$  1.3  &   20.0 $\pm$ 3.3  &   205.4  $\pm$ 11.1    &  161 $\pm$ 2.1 & 100 &       & \\
   NGC 5506     &              513.6 $\pm$ 16.2  &   38.0 $\pm$ 4.7   &  79.5 $\pm$ 6.5  &   28.3 $\pm$  3.8  &  206.8 $\pm$ 10.7 &  1614.0  $\pm$ 29.8    &  203 $\pm$ 3.0 &  98 & 42.89 & 1 \\
   NGC 5643     &              753.7 $\pm$ 19.8  &   57.4 $\pm$ 5.6   & 104.9 $\pm$ 7.4  &   61.7 $\pm$  5.7  &  224.3 $\pm$ 10.8 &  2393.0  $\pm$ 36.0    &  154 $\pm$ 2.4 &  93 & $>$40.60$^b$ & 1\\ \\
   NGC 5674     &               21.6 $\pm$ 3.3   &    3.3 $\pm$ 1.3   &   4.7 $\pm$ 1.6  &    1.2 $\pm$  0.7  &    9.3 $\pm$ 2.3  &    62.7  $\pm$  5.9    &  145 $\pm$ 4.4 & 129 & 43.18 & 1\\
   NGC 5728     &              253.1 $\pm$ 11.3  &   29.0 $\pm$ 3.8   &  50.0 $\pm$ 4.9  &   28.5 $\pm$ 3.8   &  124.9 $\pm$ 7.9  &  1154.0  $\pm$ 25.0    &  210 $\pm$ 4.1 & 155 &       & \\
   NGC 5953     &              121.4 $\pm$ 7.8   &   17.6 $\pm$ 3.5   &  37.7 $\pm$ 4.7  &    8.1 $\pm$  1.2  &   74.8 $\pm$ 6.1  &   222.2  $\pm$ 10.9    &  180 $\pm$ 13.7 &  93 &       & \\
   NGC 6221     &              201.5 $\pm$ 10.1  &   83.6 $\pm$ 6.9   & 168.1 $\pm$ 9.4  &    7.6 $\pm$  2.5  &  385.7 $\pm$ 14.6 &   249.5  $\pm$ 11.7    &  244 $\pm$ 22.0 & 111 &       & \\
   NGC 6300     &               55.3 $\pm$ 5.3   &    4.9 $\pm$ 1.6   &   6.0 $\pm$ 1.8  &    6.0 $\pm$  1.8  &   15.5 $\pm$  2.8 &   234.6  $\pm$ 11.3    &  121 $\pm$ 4.7 & 100 & 41.23 & 7\\ \\
   NGC 6890     &               64.2 $\pm$ 5.9   &   14.5 $\pm$ 2.5   &  19.2 $\pm$ 3.3  &   19.1 $\pm$  3.2  &   50.2 $\pm$  5.2 &   720.7  $\pm$ 20.1    &  204 $\pm$ 3.0 & 109 &       &  \\
   NGC 7172     &                4.8 $\pm$ 1.6   &                   &                 &                    &    3.2 $\pm$  1.1 &    15.5  $\pm$  2.8    &  132 $\pm$ 19.9 & 190 &       &  \\
   NGC 7212     &              476.3 $\pm$ 15.8  &   49.5 $\pm$ 5.1   & 100.8 $\pm$ 7.3  &   53.3 $\pm$  5.3  &  242.1 $\pm$  11.6&  2890.0  $\pm$ 39.1    &  205 $\pm$ 2.5 & 168 & 42.24 & 6 \\
   NGC 7314     &               22.0 $\pm$ 3.3   &    3.7 $\pm$ 1.5   &   8.5 $\pm$ 2.1  &    6.8 $\pm$  1.8  &   19.1 $\pm$  3.3 &   216.4  $\pm$ 10.5    &  100 $\pm$ 1.4 &  60 & 42.22 & 1 \\
   NGC 7496     &              135.1 $\pm$ 8.3   &   32.0 $\pm$ 3.9   &  58.4 $\pm$ 5.4  &                    &  150.3 $\pm$  8.9 &    96.4  $\pm$  7.3    &  205 $\pm$ 21.0   & 101 & 41.66 & 4 \\ \\
   NGC 7582     &              140.7 $\pm$ 8.5   &   49.9 $\pm$ 5.3   &  98.3 $\pm$ 7.2  &   29.8 $\pm$  3.5  &  276.0 $\pm$  12.0&   570.1  $\pm$ 17.2    &  127 $\pm$ 5.0    & 132 & 42.16 & 1 \\
   NGC 7590     &               55.1 $\pm$ 5.3   &    7.6 $\pm$ 2.4   &   9.8 $\pm$ 2.2  &    5.9 $\pm$  1.4  &   24.0 $\pm$  3.3 &   112.2  $\pm$  7.5    &   73 $\pm$ 8.7 &  99 & 40.81 & 1 \\
   NGC 7679     &              147.0 $\pm$ 8.7   &   24.0 $\pm$ 3.4   &  48.0 $\pm$ 4.8  &    8.0 $\pm$  2.8  &  114.0 $\pm$  7.8 &   154.5  $\pm$  9.9    &  176 $\pm$ 11.7 &  96 & 42.47 & 8 \\
   NGC 7682     &              262.0 $\pm$ 11.6  &   18.4 $\pm$ 3.0   &  32.6 $\pm$ 4.2  &   17.8 $\pm$  2.9  &   71.6 $\pm$  6.3 &   729.2  $\pm$ 20.0    &  139 $\pm$ 2.3 & 152 & 42.86 & 4 \\
   NGC 7743     &               83.1 $\pm$ 6.7   &    7.6 $\pm$ 2.2   &   8.5 $\pm$ 2.3  &                    &   15.7 $\pm$  2.8 &    79.2  $\pm$  6.6    &  160 $\pm$ 31.0 &  95 & 39.51 & 9 \\

\hline\hline
\end{tabular}
\end{center}

\noindent $^a$ emission line flux in units of $10^{-16}$ erg cm$^{-2} s^{-1}$.
\noindent $^b$ Compton-thick source with column density $\ge 10^{24} cm^{-2}$.
\noindent Notes:  sources of  X-ray luminosity - (1) Bassani et al. 1999; (2) Collinge \& Brandt 2000; (3) Maiolino et al. 1998;
(4) Polletta et al. 1996; (5) Iwasawa, Maloney \& Fabian 2002; (6) Risaliti et al. 2000;
(7) Matsumoto et al. 2004; (8) Della Ceca et al., 2001; (9) Terashima et al. 2002.
\end{table*}

\begin{figure}
\includegraphics[width=9cm]{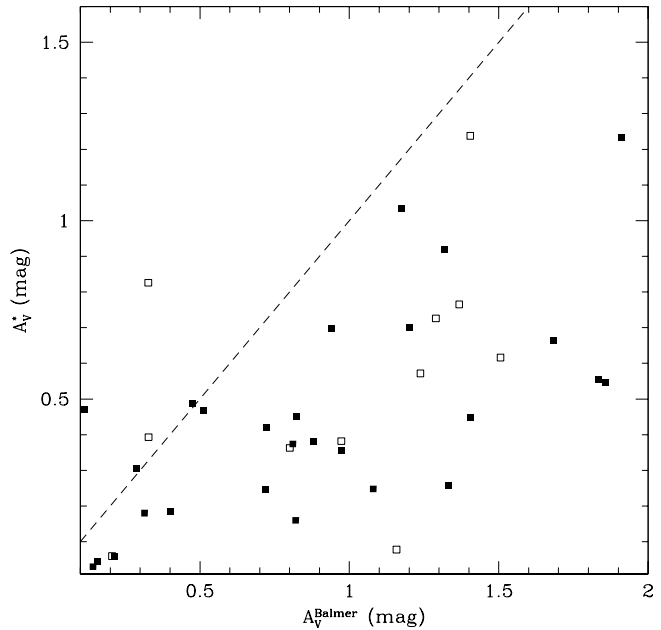}
\caption{Extinction of the stellar component ($A_{V}^{*}$) as a function
of the nebular extinction ($A_{V}^{Balmer}$) derived from the Balmer decrements
assuming Case~B recombination and the standard reddening law. Only H$\beta$
and H$\gamma$ are used to compute $A_{V}^{Balmer}$. The dashed line
represents x = y. The open symbols correspond to Seyfert 2s with direct evidence for hidden
broad-line regions (HBLR) from spectropolarimetric observations.}
\label{av}
\end{figure}

\begin{figure}
\includegraphics[width=9cm]{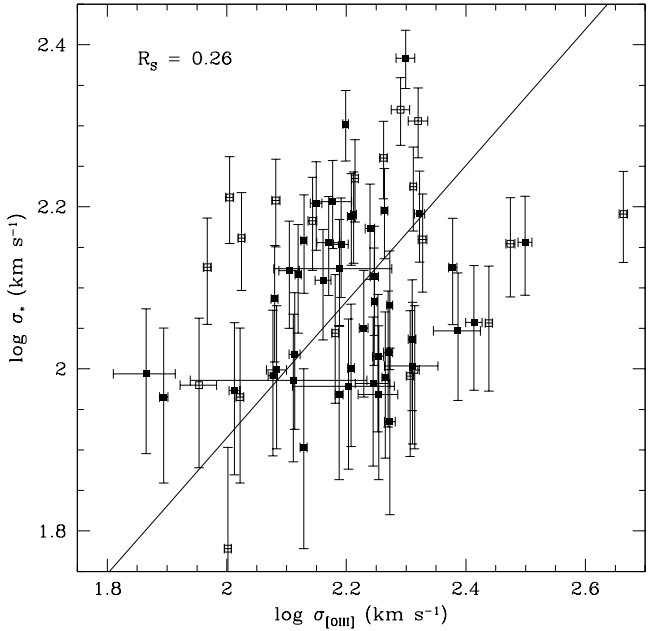}
\caption{The stellar velocity dispersion $\sigma_{*}$ derived
from the population synthesis models as a function of the velocity
width of the emission lines obtained from the [OIII]5007 line
profile.  The uncertainty in $\sigma_{*}$ is typically 20 km s$^{-1}$,
and the measurement error in $\sigma_{[OIII]}$ is estimated by a single
Gaussian fitting. The solid line shows a
ordinary least square (OLS) bisector fit to the data of slope $0.84\pm0.08$
and the Spearman rank-order correlation coefficient ($R_S$) is 0.26.
Symbols have the same meaning as in Fig. 1.}
\label{sigsig}
\end{figure}

\begin{figure}
\includegraphics[width=9cm]{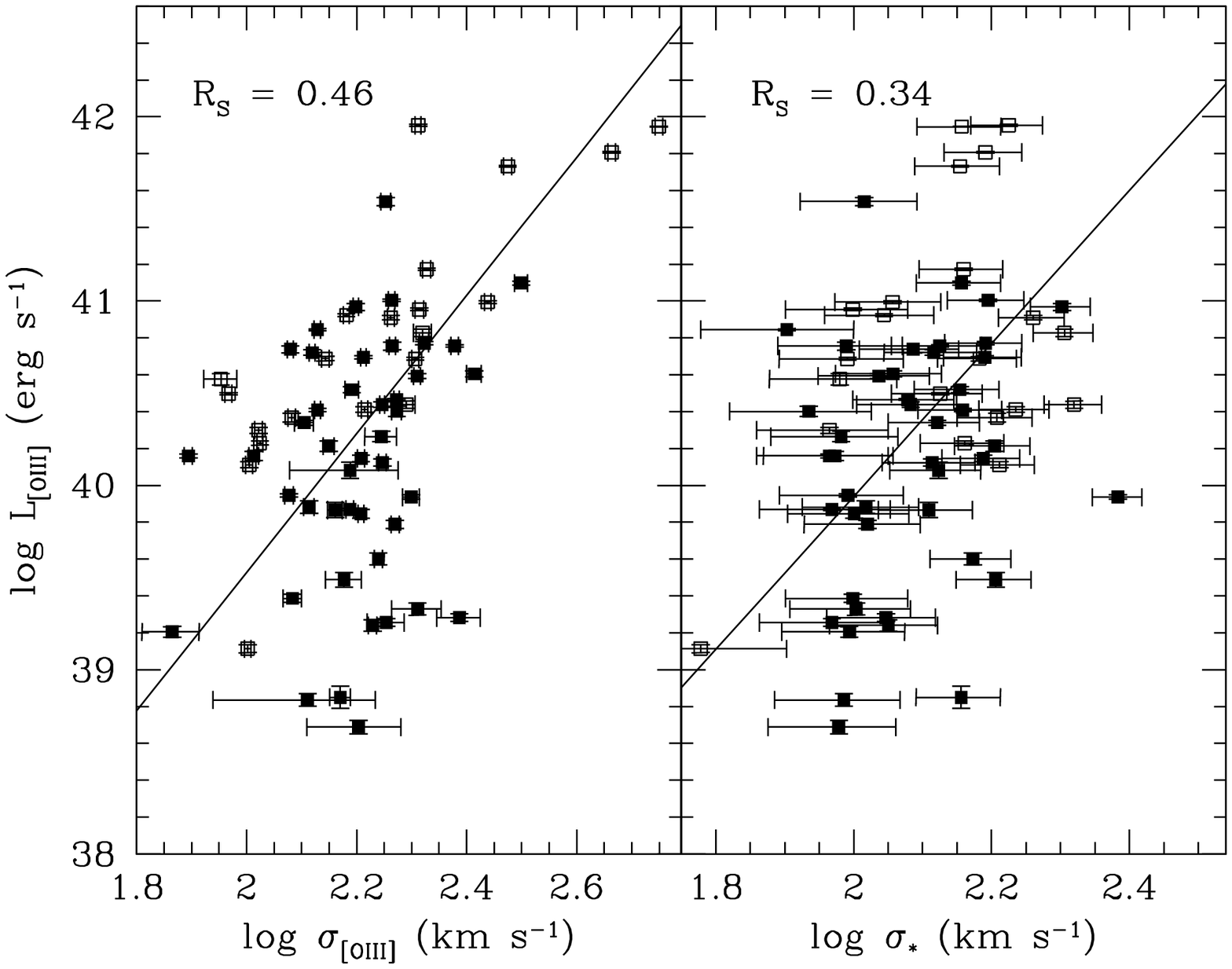}
\caption{Luminosity - velocity dispersion (L - $\sigma$) relation
for the nuclear regions of Seyfert~2 galaxies. The  correlations
of [OIII]$\lambda$5007 luminosity with the width of the nebular emission lines
($\sigma_{\rm [OIII]}$ - left panel) and with the stellar velocity
dispersion $\sigma_{*}$ (right panel) are shown.
Symbols have the same meaning as in Fig. 1, and
the lines show ordinary least square (OLS) bisector fits to the data.}
\label{lsigma}
\end{figure}

\subsection{Derived parameters}

Before deriving physical parameters from the emission line strengths
we performed the standard sanity check of computing the ratio of
[OIII]$\lambda5007$ to [OIII]$\lambda4959$ which, under all but the
most extreme physical conditions, should be equal to 3.0 (e.g. Rosa 1985).
Using only the 62 objects for which [OIII]4959 is detected with good S/N
we find a mean value of $3.00\pm0.08$.  Therefore, we conclude
that our data are not affected by unpleasant instrumental
effects (such as saturation or non-linearities) which could compromise
the reliability of the line ratios.

In order to analyze the emission line spectra, and for lack of a
better option, we will assume that the nuclear nebulae of type-2
Seyferts are similar to giant HII regions and use the
{\it classical} approach
(e.g.  Peimbert \& Torres-Peimbert, 1981).

\subsubsection{Extinction}

Assuming Case~B recombination theory and a standard reddening law
(Cardelli, Clayton \& Mathis 1989), one obtains the logarithmic reddening
correction at H$\beta$, C(H$\beta$), from the Balmer decrements (in
our data from H$\gamma$ and H$\delta$; see e.g.~Torres-Peimbert,
Peimbert \& Fierro 1989), which are

\begin{equation}
C(H\beta) = -7.41 \times \log
(\frac{F_{H\gamma}/F_{H\beta}}{I_{H\gamma}/I_{H\beta}}),  \ \ \
for \ \ H\gamma
\end{equation}

\begin{equation}
C(H\beta) = -4.95 \times \log
(\frac{F_{H\delta}/F_{H\beta}}{I_{H\delta}/I_{H\beta}}),  \ \ \
for \ \ H\delta
\end{equation}

\noindent Where $F_{H\gamma}/F_{H\beta}$, $F_{H\delta}/F_{H\beta}$
and $I_{H\gamma}/I_{H\beta}$, $I_{H\delta}/I_{H\beta}$ are the
observed and intrinsic Balmer line ratios, respectively. In this
paper we adopt the intrinsic ratios to be 0.466 and 0.256,
respectively (Osterbrock 1989). One then compares the two values
which for properly exposed and calibrated data should not differ
by more than a few percent.  In our case, even after subtraction
of the stellar component, {H$\delta$} is detected with good S/N
for 42 objects, out of which 9 have unphysical Balmer ratios (i.e.
$F_{H\delta}/F_{H\beta} >$ 0.256). {H$\gamma$ are detected with
good S/N for all objects except NGC 7172, and 14 Seyfert 2s have
unphysical Balmer ratios (i.e. $F_{H\gamma}/F_{H\beta} >$ 0.466).
For 33 objects with both good S/N and physical Balmer ratios for
H$\gamma$ and H$\delta$, } we obtain an average difference between
the two Balmer decrements of $\Delta$C(H$\beta$)$ = 0.02 \pm 0.27
(\sigma)$, which indicates that it is probably safe to use the
Balmer decrements to infer the value of the extinction for these
objects. { In the following of this paper, we will simply use
$F_{H\gamma}/F_{H\beta}$ and Equation (1) to calculate C(H$\beta$)
for Seyfert 2 galaxies and exclude NGC 7172 from further
analysis.} For objects with unphysical Balmer decrements we will
assume C(H$\beta$)=0.

We can investigate this further by comparing the nebular extinction to the
stellar extinction  (A$_{V}^{\star}$)  from our population
synthesis models.  Figure~\ref{av} shows the run of
$A_{V}^{\star}$ versus $A_{V}^{Balmer} = 0.884 \times 2.5 \times C(H\beta)$.
The nebular extinction is always larger as expected since the
stellar population is a mix of  unredenned old stars with reddened
young stars in varying proportions (Calzetti, Kinney \&
Storchi-Bergmann 1994; Mas-Hesse \& Kunth 1999).
{The open symbols represent objects where the spectropolarimetric
observations detected the polarized broad emission lines, mainly from the
compilation by Gu \& Huang (2002) plus new observations by
Lumsden, Alexander \& Hough (2004; see also Paper~II). }
In what follows we will correct our emission line
data for extinction using the Balmer decrements, but we note that none of
our conclusions actually depend on the reddening corrections.

\section{RESULTS}
\subsection{Kinematics}

 { Our stellar population synthesis code
 uses a Gaussian broadening function to match the synthetic stellar populations to the observed galaxy
 spectra, and thus provides  reliable estimates of the stellar velocity dispersion $\sigma_{*}$
 for each galaxy. We showed in Paper~II that our dispersions are in good agreement with results obtained
 using other methods and by other authors (mostly Nelson
 \& Whittle 1995; but see also Garcia-Rissmann et al., 2005).
 We have used the results of Paper~II to correct
 the emission line widths for instrumental broadening as
 $\sigma_{\rm [OIII]} = \sqrt{\sigma_{obs}^{2} -
 \sigma_{inst}^{2} }$, where $\sigma_{inst}=62~\rm km s^{-1}$.
 In what follows we will always use the
 corrected velocity dispersions in our analysis of the data.

 Figure~\ref{sigsig} presents a plot of (corrected) stellar versus nebular
 velocity dispersions. The
 solid line is the best fit to the data by means of the ordinary
 least squares (OLS) bisector method (Isobe et al. 1990),
 $\log \sigma_{*} = (0.84\pm0.08) \log \sigma_{\rm [OIII]} +
 (0.24\pm0.17)$.
 Figure ~\ref{sigsig} also quotes the Spearman rank-order correlation coefficient
 ($R_S$, Press et al. 1992) to be 0.26, and a probability of
 $P_{null} = 0.03$ for the null hypothesis of no correlation between $\sigma_{*}$
 and $\sigma_{\rm [OIII]}$.
%
%
 Our data thus confirms the findings of Nelson \&
 Whittle (1996) and Jimenez-Benito et al. (2000) of a correlation
 between nebular and stellar velocity
 dispersions,  albeit with a very significant scatter.
 In general we expect the emission line-profile widths to be due
 to the combination of hydrodynamical effects (winds, jets) and the motions of individual
 line emitting clouds in the gravitational potential of the
 nuclear regions. Thus, if the nebular and the stellar components
 feel the same gravitational field, then the emission lines should be
 systematically broader than the stellar lines. 
 In fact, 18 of our Seyfert 2's show stellar velocity
 dispersions larger than the nebular ones, although only for 7 of these,
 (ESO 434-G40, NGC 788, NGC 2110, NGC 3035, NGC 4903, NGC 7172 and NGC 7590),
 the differences are larger than the typical errors in
 stellar velocity dispersion (20km/s). For NGC 7172, the difference is  due to
 the low signal-to-noise of the spectrum, while 4 of the remaining 6 have clearly non-Gaussian emission lines with
 real widths significantly broader than the single Gaussian fits. Thus, for only two objects in our sample - NGC 788 and NGC 3035 - the stellar lines
 appear to be significantly broader than the nebular lines.
 
\begin{figure}
\includegraphics[width=9cm]{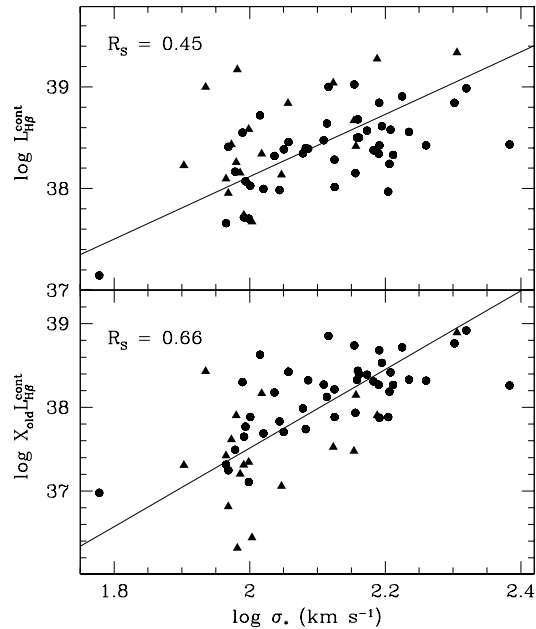}
\caption{The relationship between monochromatic continuum
luminosity ($L_{H\beta}^{cont}$) and stellar velocity dispersion
for Seyfert~2 galaxies. Upper panel: the continuum luminosity is
estimated from the total 4861 \AA \ continuum emission; Lower
panel: Continuum luminosity is  estimated using only the
luminosity from the old contribution.  Squares (triangles)
indicate Seyfert 2s with W$_K < 10$\AA ( ($\ge$ 10 \AA).  The
y-axis are in units of $erg s^{-1} \AA ^{-1}$. } \label{lsigmab}
\end{figure}

\begin{figure}
\includegraphics[width=9cm]{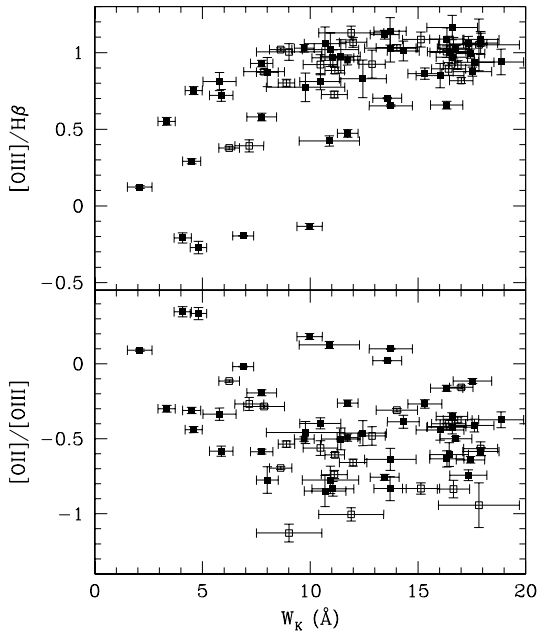}
\caption{Relation between gaseous excitation (Upper: [OIII]/H$\beta$
 and Lower: [OII]/[OIII]) and equivalent width of
 CaII K (a good indicator of the age of nuclear stellar populations)
 from population synthesis models.  Symbols have the same meaning as in Fig. 1.
 Objects with HBLRs (open symbols) are not
 systematically different from the rest of the sample in these plots.}
\label{ex}
\end{figure}

 It is well known that [OIII] luminosity is a good tracer of the
strength of the nuclear activity (Mulchaey et al. 1994; Kauffmann
et al. 2003b), Figure~\ref{lsigma} shows the relations between
$L_{\rm [OIII]}$ and the velocity dispersions of the nebular and
stellar components. We remark that, although our
extinction-corrected [OIII] luminosities have been observed and
reduced in a homogeneous way, there is still an intrinsic source
of scatter arising from the fact that, as a function of distance,
our spectrograph slit encompasses different fractions of the total
nuclear emission for different objects.  The lines show ordinary
least square (OLS) bisector fits to the data of slopes $3.8\pm0.4$
if we use $\sigma_{\rm [OIII]}$, and $ 4.1\pm1.0$ for
$\sigma_{*}$.  The Spearman rank-order correlation coefficients,
$R_S$, are  0.46 and 0.34 and $P_{null}$  $< 10^{-4}$ and 0.007,
respectively. Two features of these plots will be important in the
discussion: (1)  the correlation is clearly better for the nebular
velocity dispersions, and (2) the objects with indications of
hidden broad line regions lie predominantly above the best fit
line in the $\rm L_{[OIII]} - \sigma_{\rm [OIII]}$ plot.

  The Faber-Jackson relation (Faber \& Jackson 1976) is the
correlation between  bulge (continuum) luminosity  ($L_{Bulge}$)
and  central (stellar) velocity dispersion.  Since we do not have
$L_{Bulge}$, we will use the monochromatic continuum luminosity at
$\lambda=4861$\AA, $L_{H\beta}^{{cont}}$, as a proxy.
Figure~\ref{lsigmab} (upper panel) shows the relation between the
$L_{H\beta}^{{cont}}$ and $\sigma_\star$.  We have  used different
symbols to parametrize the nuclear stellar populations by means of
W$_{\rm K}$  -- the equivalent width of the  CaII K absorption
line -- which is a very  good indicator of the mean age of the
stellar population (e.g. Cid Fernandes \etal 2001;  and Paper II).
Thus, triangles represent young objects ( W$_{\rm K}<10$\AA)  and
squares galaxies with W$_{\rm K} \geq 10$\AA\   corresponding to
objects whose continuum is dominated by old populations. The plot
is quite scattered, but shows interestingly (top panel)  that
 Seyfert 2s with  $W_{\rm K}<10$\AA\ lie systematically above the
 best fit line, indicating lower M/L ratios.  If we use the
 continuum  from the old populations only,
 $L_{H\beta}^{cont}\times x_{old}$, where $x_{old}$  
 is the fraction of stars older than  10$^9$ yrs from our stellar population synthesis models (Paper II),   
  the scatter is significantly reduced  and the `young' objects are more
 symmetrically distributed around the best fit line (lower panel).  Furthermore,
 the slope  changes from 3.0 $\pm$ 0.5 (top panel) to 4.7 $\pm$
 0.5 (lower panel), closer to the actual
 Faber-Jackson relation.  We confirm, therefore, that starburst
 activity  distorts the Faber-Jackson relation of Seyfert galaxies
 as originally proposed by Nelson \& Whittle (1996) and Gu et al.
 (1998).

\subsection{Physical conditions of the nebular gas}

 The excitation of the nebular gas provides us with a further test
 of the link between the source of ionizing photons and the
 nuclear stellar populations.  Figure~\ref{ex} plots two
 indicators of nebular excitation ([OIII]$\lambda$5007/H$\beta$
 and [OII]$\lambda$3727/[OIII]$\lambda$5007) as a function of
 $W_{K}$ described above. {The Spearman tests gives R$_S$  0.22
 and -0.58 (P$_{null}$ 0.09 and $< 10^{-4}$), respectively. The
 trend of younger nuclei having lower excitation and vice versa
 is clearly seen in [OII]/[OIII], but the separation is less clear
 in [OIII]/H$\beta$.} This  is probably due to  circum-nuclear
 starburst activity that powers a substantial fraction of the
 H$\beta$ emission, but little of the [OIII] and HeII fluxes in
 nuclei  with low $W_K$. This would result in a dilution of the
 line ratios relative to the more extreme values attained by AGN
 without conspicuous star-formation (large $W_K$; c.f. Cid
 Fernandes et al.\, 2001). The fact that  [OII]/[OIII]  shows a
 clear trend suggests that [OII] has a significant contribution
 from circumstellar HII regions. A similar result was found in
 composite low luminosity QSO spectra from the 2dF/6dF surveys
 (Croom et al 2002; but see Ho 2005).

 There is also a relation between the nuclear population
 ages and H$\beta$ luminosity, as shown in Figure~\ref{ageL}, which
 also shows that younger objects are more luminous in H$\beta$.  It can
 be verified using the data in Table~1 that these trends hold even at
 fixed velocity dispersions. That is, for a given value of $\sigma_\star$,
 younger objects tend to be systematically above the correlation line
 shown in Figure~\ref{lsigma}, thus indicating that trends in
 luminosity with age reflect real changes in the stellar mix.

\begin{figure*}
\includegraphics[width=15cm]{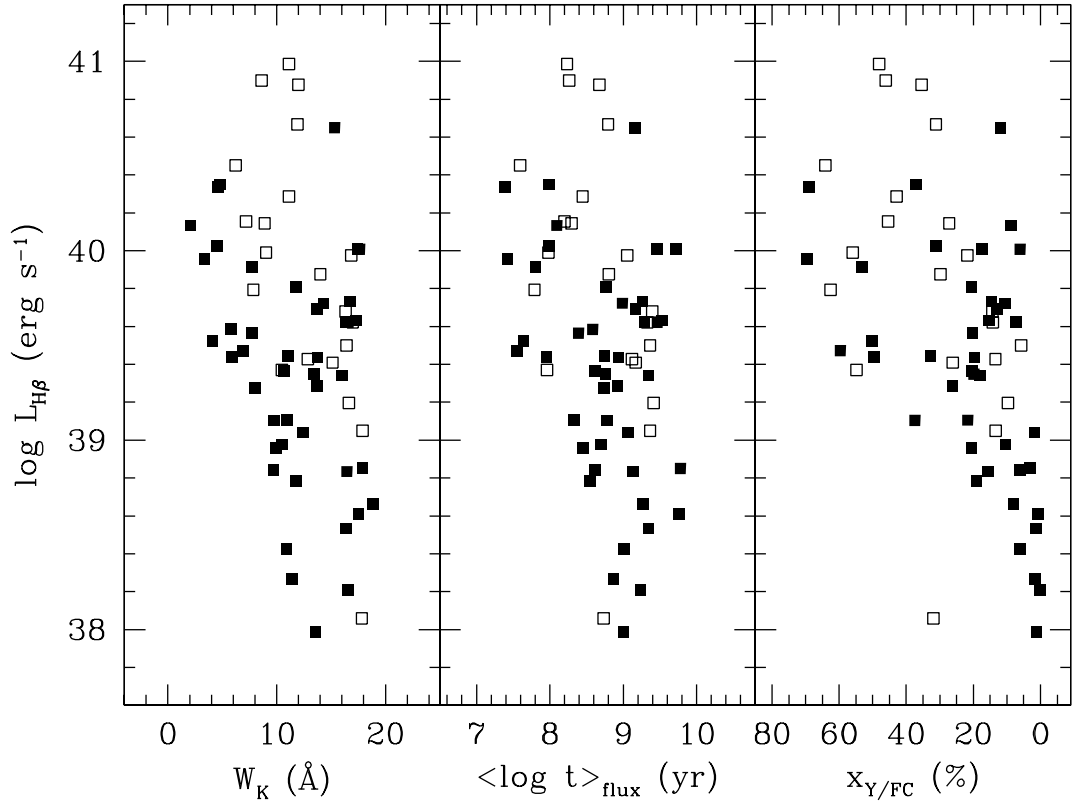}
\caption{Relation of H$\beta$ luminosity as a function of 3
indicators of the mean age of the nuclear stellar population:
$W_{K}$, mean age (log~t), and fraction of the total continuum
emitted by young stars to the featureless (non thermal)
continuum $x_{Y/FC}$. Symbols have the same meaning as in Fig. 1.}
\label{ageL}
\end{figure*}

\section{Discussion}

\subsection{Starbursts as sources of ionizing photons}

 The young stellar components detected in many Seyfert 2s are bound
 to contribute to the ionization of the surrounding gas. One should
 thus expect a proportionality between the optical continua associated
 with these components and the H$\beta$ luminosity. This is confirmed in
 Figure~\ref{nLyc}, where we plot the observed H$\beta$ emission
 line fluxes as functions of the ionizing stellar and featureless
 continua at 4861\AA, both corrected for the extinction derived from
 the synthesis models of Paper~II (see also Cid Fernandes \etal 2001).
 The lines correspond to constant equivalent widths of 10\AA, 100\AA, and
 1000\AA\ in each plot. From the left to the right panels of
 Figure 7, the Spearman ranks  are 0.60, 0.71 and 0.84 respectively
 showing that the correlation gets much better when we only use the
 optical flux associated with the putative ionizing continua ($\rm
 x_{Y/FC} f_{H\beta}^{cont}$) instead of the total stellar continuum
 light.

\begin{figure*}
\includegraphics[width=15cm]{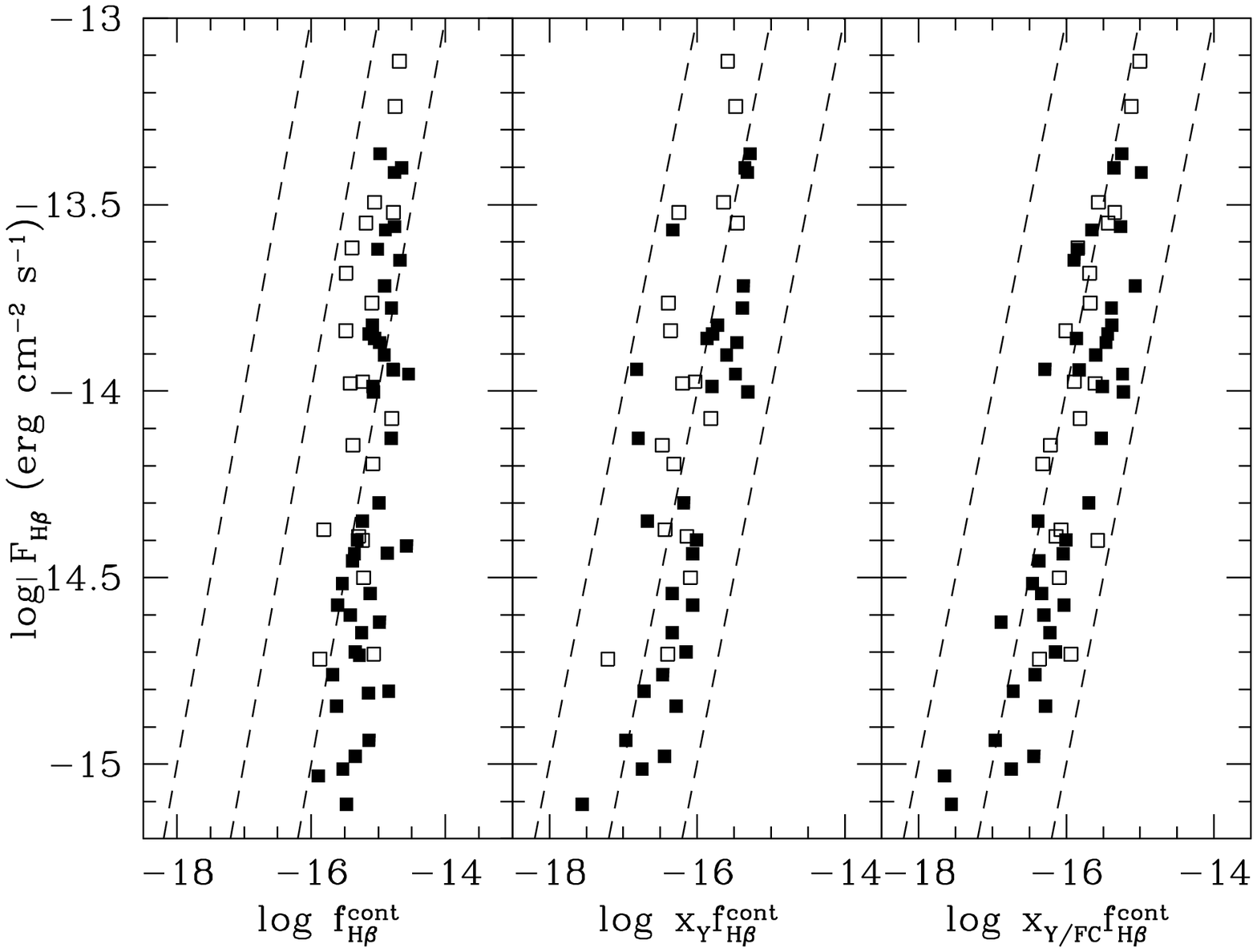}
\caption{Relation between the continuum at H$\beta$ and H$\beta$
emission line fluxes. The panels differ in which component of the
continuum is used in the abscissa. (left) total observed
continuum, (middle) only the continuum due to young components;
(right) the continuum from young and FC components. All continua
are given in units of erg$\,$s$^{-1}$cm$^{-2} \AA^{-1}$. The dashed lines
represent, from left to right, loci of constant equivalent widths of 10, 100
and 1000 \AA . The objects with HBLRs (open symbols) are not segregated
in these plots.}
\label{nLyc}
\end{figure*}

 We can also use our stellar population synthesis
 results to predict the H$\beta$ luminosities from the number of
 Lyman continuum ionizing photons, $Q_{H}$, emitted by each stellar component 
 using the theoretical values given by Bruzual \& Charlot (2003) for each stellar base.
 Since the  flux of Lyman continuum photons, $Q_H$, changes by
 more than one order of magnitude during the first few Myrs of a
 starburst, whereas the optical spectrum changes relatively
 little, the young (5~Myr) component of our models represents the
 optical light very well,  but the corresponding $Q_H$ may be
 wrong by  factors of 10 or more. Therefore we have calculated an
 extended grid of  models adding two additional young components
 of 1~Myr and 3~Myr.

 Figure \ref{nLyc2}(a) compares the observed $L_{H\beta}$ with the
 values predicted by our models using {\it  a single age}  of
 5~Myr for  the young stellar component, $L^\star_{H\beta}$. No
 clear correlation is seen, and in fact the observed luminosities
 are systematically larger than the predicted ones (the one-to-one
 correlation is indicated by the dashed line).  As discussed
 above, $L^\star_{H\beta}$ depends strongly on the ages of the
 youngest components included in the synthesis models. This is
 illustrated in Figure \ref{nLyc2}(c) which shows the comparison
 between the predicted and observed luminosities obtained using
 the extended models.

 The predicted luminosities, however, (Figure \ref{nLyc2}c) still
 do not include the FC component in the computation of $Q_H$.
 Figure~\ref{nLyc} demonstrates that, regardless of its true
 nature (young stars or reflected light from a hidden AGN, 
 Storchi-Bergmann et al. 2000; Gonzalez Delgado et al.
 2001 and Cid Fernandes et al. 2001), the FC component
 is associated with ionizing photons, so it must be included in
 the calculation of $Q_H$, the question is how.  Here we have
 converted the FC continuum luminosities from the  models
 ($x_{FC}L^{cont}_{H\beta}$) to  emission-line luminosities (
 $L^{FC}_{H\beta}$)  by assuming that photo-ionization {\it by the FC
 component alone} would result in an EW(H$\beta$) of 100\AA, 
 typical value for both
 starburst and Seyfert nuclei (Figure~\ref{nLyc}).  This estimate
 is then added to $L^\star_{H\beta}$ to compute the total
 predicted H$\beta$ luminosity. The results are shown in panels
 (b) and (d) for the two stellar bases considered above. In both
 cases we obtain very good correlations, in particular when we use
 multiple young stellar components (d) for which the points are
 very well fitted by a line of slope one.

\begin{figure*}
\includegraphics[width=15cm]{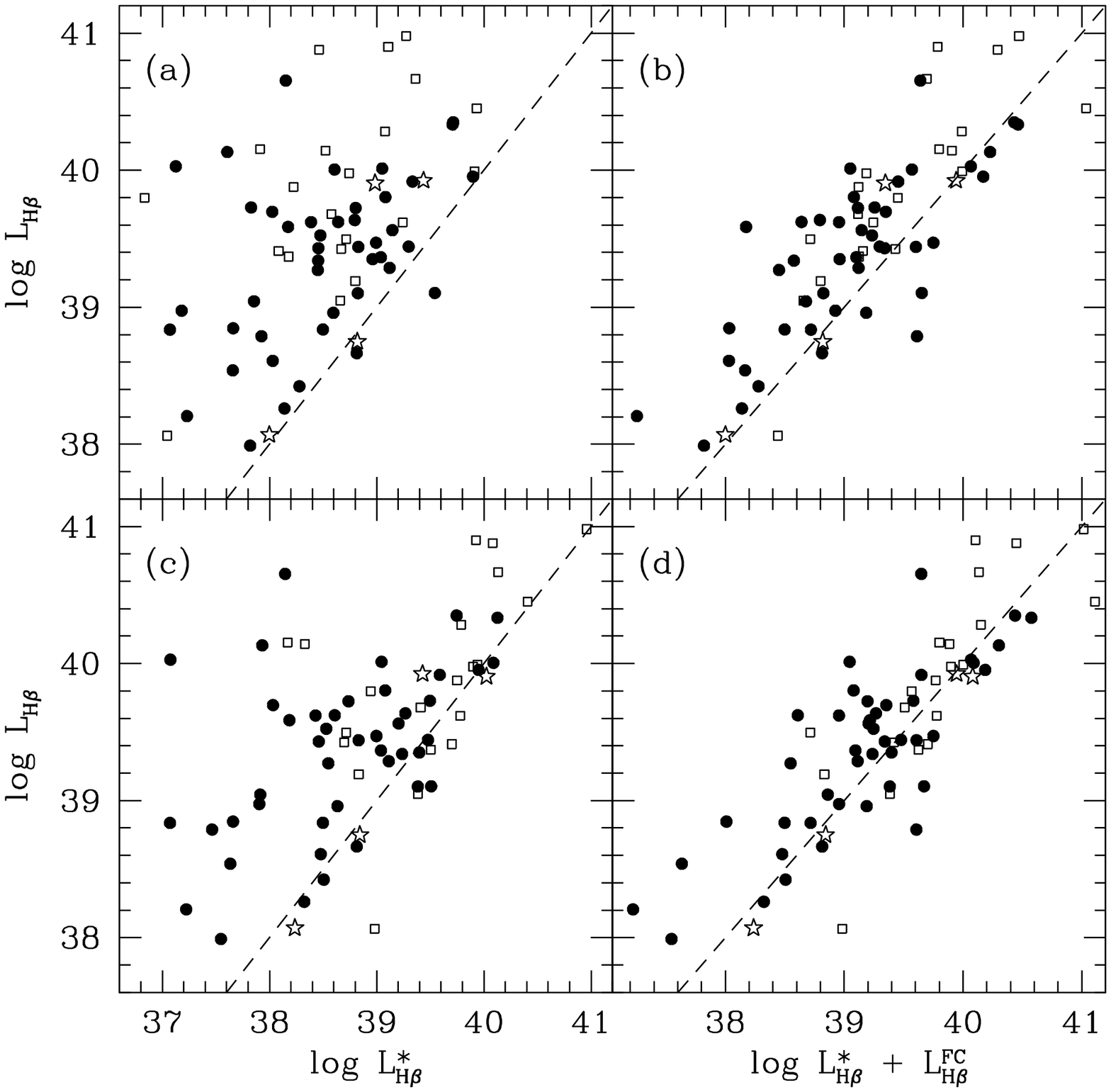}
\caption{
Predicted versus observed H$\beta$ luminosities (all in
erg$\,$s$^{-1}$). $L^\star_{H\beta} = 4.76 \times 10^{-13} Q^\star_H$
is the H$\beta$ luminosity powered exclusively by the stellar
components in the base. See text for the FC contribution to the rate of
ionizing photons.
 (a) $L^\star_{H\beta}$
versus $L^{obs}_{H\beta}$ for the 5 Myr stellar base used in this paper.
(b) Same as (a), but adding $L^{FC}_{H\beta}$ to $L^\star_{H\beta}$.
(c) $L^\star_{H\beta}$ obtained with a
larger stellar base (1, 3 and 5 Myr). (d) As (b), but for the enlarged base used in (c).
As usual HBLR objects are plotted with open-squares.
The 4 starburst galaxies in Joguet's sample are shown as open stars.}
\label{nLyc2}
\end{figure*}

 In order to check the consistency of the models we have also
plotted in Figure~\ref{nLyc2} the predictions for the 4 objects in
Joguet's original sample that are pure starbursts.  These objects,
shown as stars, lie exactly where the predicted and observed
H$\beta$ luminosities are equal, thus validating the internal
consistency of our models.  It is interesting to notice that,
while for  3 of these objects the synthesis yields a relatively
small FC components (due to the younger starbursts), 
for one, NGC 3256, the inclusion of the FC
increases the predicted H$\beta$ luminosity by a factor of $\sim
3$, which illustrates the ambiguity of the FC discussed above and
in Paper II.

We can now use our models to calculate the fraction of the
ionization flux that is produced only by the starburst components,
$f_{SB} = L^{{obs}}_{H\beta}/L^{pred}_{H\beta}$, shown in
Figure~\ref{ragn}, where for $L^{pred}_{H\beta}$ we have summed
all ionizing photons supplied by the stellar components in the
base except for the FC component, though it could be due to a 
young dusty starburst. The median fraction of starburst
ionizing power is 65\%, while for only 20\% of the sample the
starburst contribution is less than 10\% of the total power
required to ionize the gas.  These figures are lower-limits since
with our data we cannot distinguish AGN from dusty starbursts,
but, even if all the FC came from the AGN, the distribution is
still consistent with 50\% of the objects being fully powered by
starbursts.

\begin{figure}
\includegraphics[width=9cm]{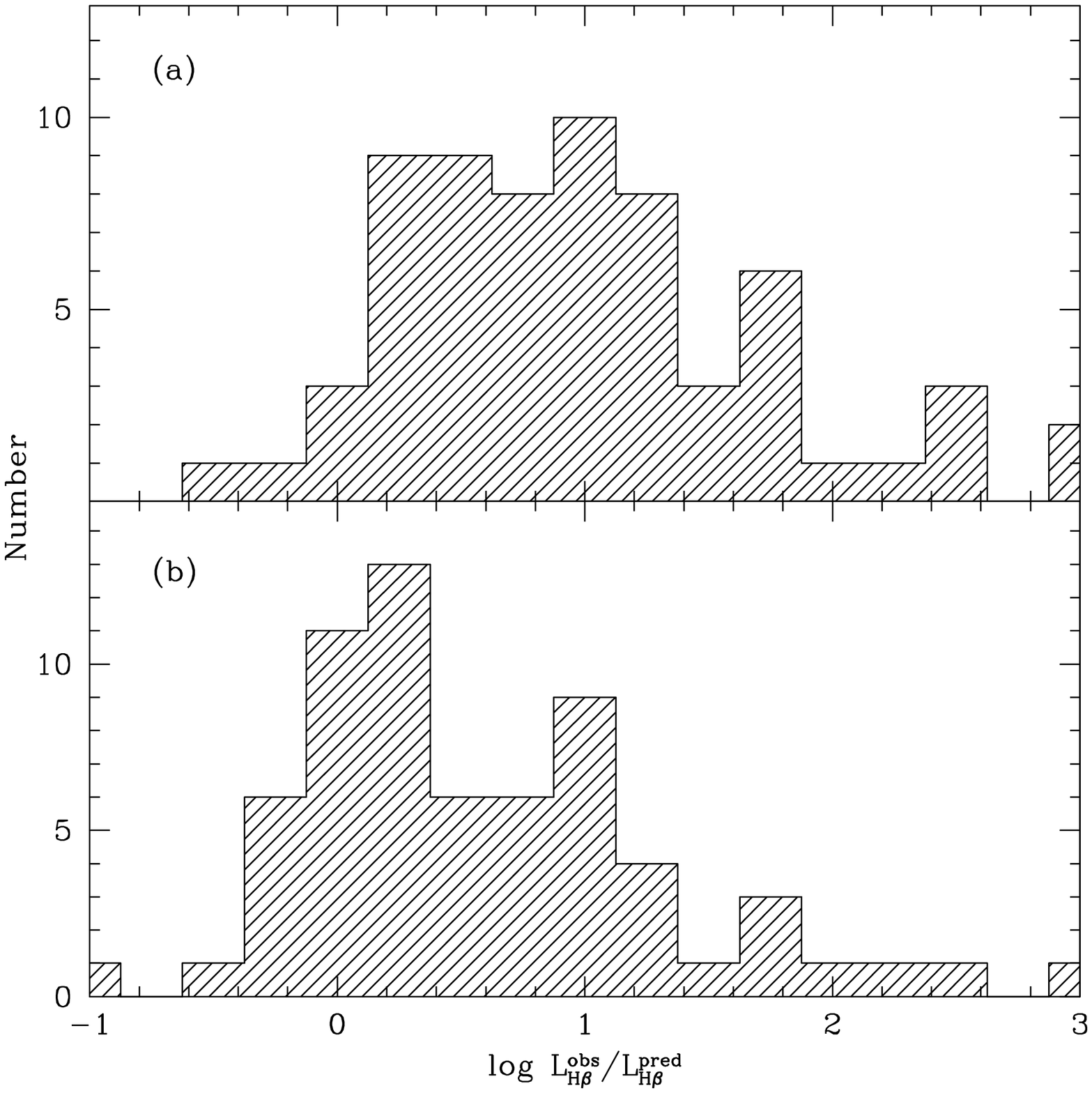}
\caption{Histogram distribution of the ratio of the observed $L_{H\beta}$
 to the values predicted by the stellar components in the base.
 (a): only one at 5 Myr; (b): 1, 3 and 5 Myr models.}
\label{ragn}
\end{figure}

To summarize, besides the presence of weak (scattered) broad-line
components in some nuclei, our optical spectra do not provide us
with reliable direct  tracers of the properties, nor even of the
existence, of the hidden AGN in the nuclear regions of nearby
low-luminosity Seyfert~2 galaxies.

\subsection{Starbursts versus Monsters}

Within the ambiguities of  the FC discussed above, the parameter
$f_{AGN}=1-f_{SB}$ should provide us with at least  a hint of the
ionizing power of  the elusive hidden AGN.   Figure~\ref{fagnis}
plots the relation between  $f_{AGN}$  and nebular excitation. In
order not to crowd the plots we have left out the HeII lines for
which we do not see any significant correlations, and which may be
compromised by Wolf-Rayet features.
\begin{figure}
\includegraphics[width=9cm]{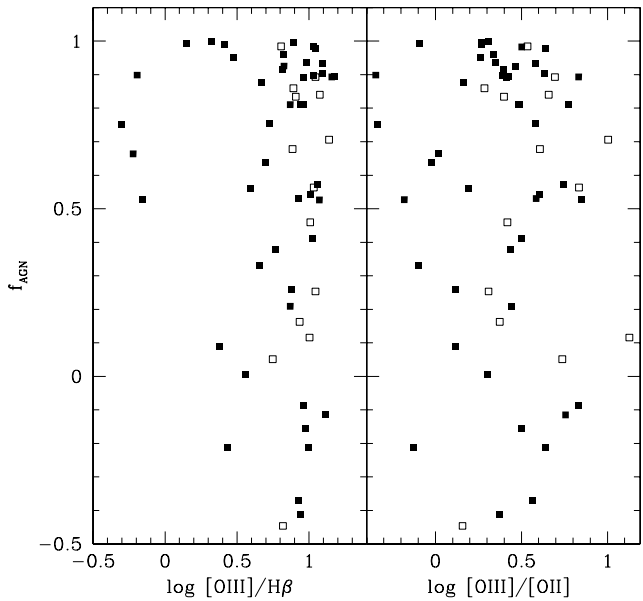}
\caption{ Relationships between  the maximum Lyman ionization power produced by
a hidden AGN component, $f_{AGN}$, and two indicators
of the excitation of the nebular gas.  As before, open squares correspond to objects with HBLRs.}
\label{fagnis}
\end{figure}

While these plots show an intriguing structure, there is no
obvious correlation between AGN fraction and excitation. Moreover,
objects with putative HBLRs (open symbols) do not seem to stand
out in these plots either. Since excitation is one of the
classification criteria for Seyferts, these plots indicate that
$f_{AGN}$ is not a very good starburst/monster discriminator.
Another  potentially powerful  indicator of accretion power is
hard X-ray luminosity. Figure~\ref{lsX} (left panel) shows the
relation between $f_{AGN}$ and absorption-corrected  hard X-ray
luminosity.  The lower-limits (arrows) correspond to Compton
self-absorbed objects (hydrogen column densities  $\rm
>10^{24}cm^{-2}$) for which only reflected X-rays are observed.
The right panel shows $L_X$ as a function of [OIII]/[OII].
\begin{figure}
\includegraphics[width=9cm]{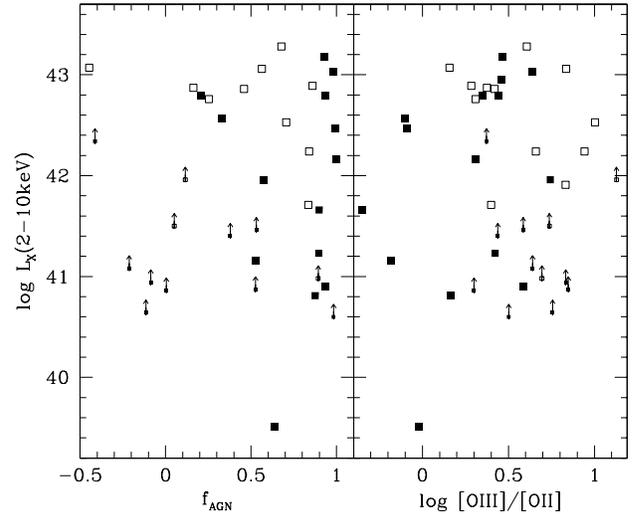}
\caption{Relationships between $f_{AGN}$, nebular excitation and
hard X-ray luminosity.  Arrows show Compton self-absorbed galaxies
for which the X-ray luminosities are lower-limits. Galaxies with
direct indications of a hidden BLRs are plotted as open squares.}
\label{lsX}
\end{figure}

There is no  correlation between hard L$_{X}$  and  AGN fraction,
but there is a clear trend of nebular excitation being higher for
more X-ray luminous objects.  These plots also show very clearly
that nuclei with HBLRs have high X-ray luminosities, although the
interpretation of this result is not straightforward since there
may be a substantial contribution to the X-rays  from off-nuclear
sources (Jimenez-Bail\'on et al., 2003; Levenson et al., 2005). It
is therefore relevant to examine the relation between X-ray
luminosity and the global parameters of the sample:  velocity
dispersion and H$\beta$ luminosity. These relations are presented
in  Figure~\ref{XLsigma} where we see a weak trend of $L_{X}$ with
$\sigma_{\rm [OIII]}$, but a significant correlation with
L(H$\beta$) which is not a distance effect because it is also
present in the fluxes. This is shown in Figure~\ref{FluxFlux}
where for clarity we have omitted the lower-limits.  The line
shows the OLS fit to the data of slope $1.1\pm 0.2$ ($R_S=0.19$)
so, with large scatter, the X-ray power is roughly proportional to
the Lyman ionizing power. Since we have no control over the
spatial regions covered by the two data sets, we should be very
careful about how to interpret this plot. Taken at face value it
means that the bulk of the X-rays are produced by the same sources
that  photoionize the gas.  This is consistent with the fact that the few objects with
[OIII]/H$\beta < 3$ (cf. Figure~10) lie  below the best
fitting line. 

\begin{figure}
\includegraphics[width=9cm]{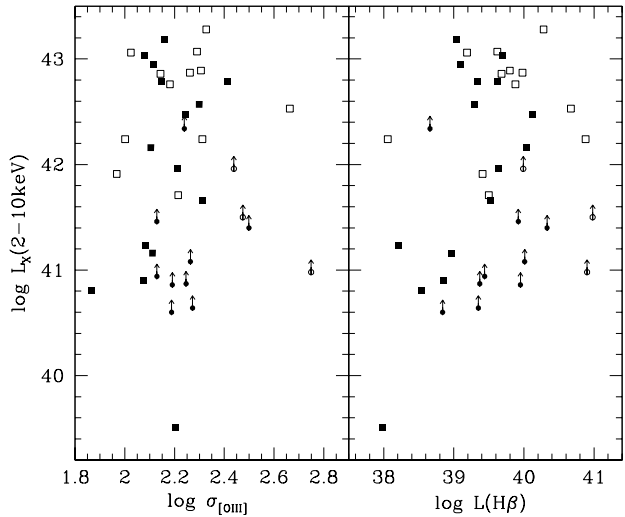}
\caption{Correlations between (2-10keV) X-ray luminosity, nebular velocity
dispersion ($\sigma_{\rm [OIII]}$), and H$\beta$ luminosity.  As before, arrows
correspond to Compton self-absorbed lower limits, and open symbols to
galaxies with hidden BLRs.}
\label{XLsigma}
\end{figure}

\begin{figure}
\includegraphics[width=7cm]{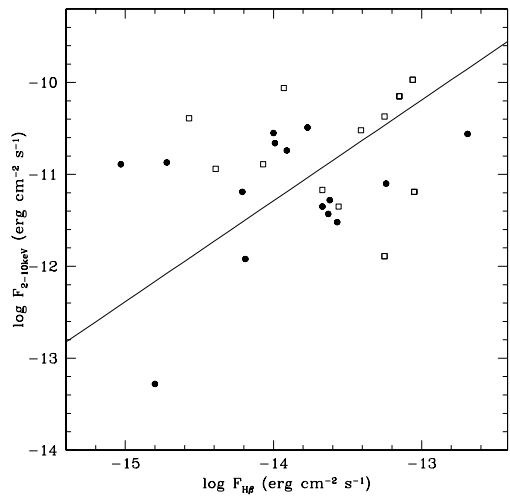}
\caption{ 2-10keV X-ray flux against H$\beta$ flux.  We have omitted the Compton self-absorbed sources for clarity.  The line shows the best fitting line of slope unity.}
\label{FluxFlux}
\end{figure}

Since even using deep high resolution X-ray (Chandra) and UV (HST)
images it remains extremely difficult to see the AGN in
nearby Seyfert~2's
 ( e.g.. Jim\'enez-Bail\'on et al. 2003; Levenson et al. 2005),  and since even when it is seen, there is still a significant contribution from circum-nuclear sources, the only way to answer the {\it Starburst or Monsters} question is to have deep high spatial resolution images of a substantial sample of nearby Seyfert~2 galaxies over a wide wavelength range. Until this is available, the interpretation of integrated observations will be plagued with the same uncertainties that we have confronted in the present work.

The INTEGRAL satellite has recently unveiled a putative new class
of powerful, highly obscured, X-ray sources associated with
massive stars: High Mass X-ray Binaries (HMXBs; Chaty and
Filliatre, 2005).  If HMXB's represent a stage in the evolution of
high-mass binaries, they could be very abundant in the
circumnuclear starbursts presented in many Seyfert 2s, and thus
potentially explain their X-ray morphologies.


\section{Conclusions}

We have presented a study of pure emission line spectra of a
sample of 65 Seyfert 2 galaxies  obtained by subtracting the
synthetic stellar component from the observed spectra. Our main
results are summarized as follows.

(i) For most Seyfert 2 galaxies, the gaseous nebular extinction
deduced from the Balmer decrements appears to be systematically larger than
the stellar extinction.

(ii) We compare the velocity dispersion of the stars and ionized gas,
and confirm the existence of correlation between stellar and nebular velocity dispersion.

(iii) Seyfert 2 nuclei follow a Faber-Jackson like correlation between
nuclear continuum luminosity and stellar velocity dispersion, but
nuclei which also harbor star-formation deviate systematically from
this relation. Removing the young stars from the continuum luminosity
(using the synthesis results) strengthens the correlation
substantially, and results in a slope more compatible with the actual
Faber-Jackson relation.

(iv) There is a relation between ionized gas velocity dispersion
and emission-line  luminosities. We also find that Seyfert 2s with
indication of broad lines either from the pure emission line spectrum
or spectropolarimetric observations show larger [OIII]
luminosities.

(v) We confirm our earlier inference that photoionization by young
stars can explain a substantial fraction of the nuclear emission-line
luminosities of Seyfert~2 nuclei, but neither the parameters derived from our optical spectra,
nor the hard X-ray luminosities provide accurate indicators of the AGN contribution.

(vi) Albeit with substantial scatter, there is a correlation between hard (2-20keV) X-ray luminosity and H$\beta$ luminosity for our sample of low-luminosity AGN.
The recent discovery of a new population of highly obscured luminous X-ray sources associated with high mass stars
 may yield some new clues to the interpretation of the properties of low-luminosity Seyfert galaxies. The off-nuclear X-ray sources detected by Chandra in some of these galaxies may, in the end, be powered by the same starbursts that are producing the bulk of the optical emission.

\section*{ACKNOWLEDGMENTS}
 The authors are very grateful to the anonymous referee for his/her
 careful reading of the manuscript and instructive comments
 which significantly improved the content of the paper.
QG thanks the hospitality of UFSC and ESO-Santiago and the support
from CNPq and the National Natural Science Foundation of China
under grants 10103001 and 10221001 and the National Key Basic
Research Science Foundation (NKBRSG19990754). Partial support from
CNPq and Instituto do Mil\^enio are also acknowledged. JM wishes
to thank the hospitality of Nanjing University during a visit
where the first manuscript of this paper was completed. ET and RT
acknowledge support by the Mexican research council (CONACYT)
under grants 32186-E and 40018 and thank the hospitality of the
IAP and the financial support from a PICS grant for a visit to
Paris to facilitate part of this work.

\end{document}